\documentclass[journal,10pt]{IEEEtran}

\usepackage{multirow}
\usepackage{graphicx}
\usepackage{amsmath,amssymb}
\usepackage{multirow}
\usepackage{graphicx}
\usepackage{cite}
\usepackage{amsmath,amssymb,cite,graphicx,epsf,color,url,booktabs,tabularx}
\usepackage{float}

\newcommand{\bm}[1]{\mbox{\boldmath{$#1$}}}

\newtheorem{Thm}{Theorem}

\begin{document}

\title{Joint User Identification and Channel Estimation Over Rician Fading Channels}

\author{Liang Wu, {\it Member, IEEE}, Zaichen Zhang, {\it Senior Member, IEEE}, Jian Dang, {\it Member, IEEE}, Yongpeng Wu, {\it Senior Member, IEEE}, Huaping Liu, {\it Senior Member, IEEE}, and Jiangzhou Wang, {\it Fellow, IEEE}

\thanks{Copyright (c) 2015 IEEE. Personal use of this material is permitted. However, permission to use this material for any other purposes must be obtained from the IEEE by sending a request to pubs-permissions@ieee.org.}
\thanks{L. Wu, Z. Zhang and J. Dang are with National Mobile Communications Research Laboratory, Southeast University, Nanjing 210096, China (e-mail: wuliang@seu.edu.cn; zczhang@seu.edu.cn; dangjian@seu.edu.cn).}
\thanks{Y. Wu is with the Department of Electronic Engineering, Shanghai Jiao Tong University, China, Minhang 200240, China (e-mail: yongpeng.wu@sjtu.edu.cn).}
\thanks{H. Liu is with the School of Electrical Engineering and Computer
Science, Oregon State University, Corvallis, Oregon 97331, USA (e-mail: hliu@eecs.oregonstate.edu).}
\thanks{J. Wang is with School of Engineering and Digital Arts, University of Kent,
Canterbury, CT2 7NT United Kingdom (e-mail: j.z.wang@kent.ac.uk)}
}

{}

\maketitle

\begin{abstract}
This paper considers crowded massive multiple input multiple output (MIMO) communications over a Rician fading channel, where the number of users is much greater than the number of available pilot sequences. A joint user identification and line-of-sight (LOS) component derivation algorithm is proposed without requiring a threshold. Based on the derived LOS component, we design a LOS-only channel estimator and an updated channel estimator.
\end{abstract}

\begin{IEEEkeywords}
Channel estimation, crowded massive multiple input multiple output (MIMO), Rician fading, user identification.
\end{IEEEkeywords}

\IEEEpeerreviewmaketitle

\vspace*{-0.2in}
\section{Introduction}
Massive multiple input multiple output (MIMO) is one of the key technologies in the fifth generation (5G) wireless communications \cite{R1}. The advantages of massive MIMO include high channel capacity, high energy efficiency, robustness to fast fading, etc. In a massive MIMO system, the base station (BS) is equipped with a massive number of antennas, and tens of users can communicate with the BS simultaneously through spatial division \cite{R7}.
Recently, crowded massive MIMO communication has been gaining increasing attention \cite{R11,R12,Rf6}. In crowded massive MIMO communications, a large number of users in a limited spatial region attempt to communicate with the BS equipped with massive antennas. Typical application scenarios include shopping mall, stadium, or open air festival \cite{R11}. In such systems, the number of users is greater than the number of available pilot sequences, and it is impossible to assign a unique orthogonal pilot to each user.
Therefore, random pilot allocation was proposed in \cite{R11,R12}. However, this also causes a critical issue in such massive MIMO systems - pilot contamination. A strongest-user collision resolution protocol was proposed in \cite{R11} to address the pilot contamination in massive MIMO systems. In the random-pilot-allocation scheme, a unique pilot-hopping pattern, which is determined by the pilots over multiple transmission slots, is assigned to each user, and the BS identifies the active users via detecting the pilot-hopping pattern aided by a predefined threshold \cite{R12}. If the threshold is not set properly, user identification will not be accurate. In the performance analysis of \cite{R12}, perfect user identification was assumed. Besides, if some of the active users employ the same pilot sequence in the same transmit slot, pilot contamination will be severe.

Rayleigh fading channel was considered in \cite{R11,R12}. In Rayleigh fading channels, there is no direct line-of-sight (LOS) path between a transmitter and a receiver. Rayleigh fading channel model approximates well for the channel in urban and macro-cellular environments \cite{Rf1}. In the massive connection scenario, \cite{Rf7,Rf8,RfCao} modelled the sparsity as the prior distribution of the channel, and applied an approximate message passing (AMP) algorithm to detect active users in Rayleigh fading channels. In open areas and micro-cellular environments, an LOS component is generally present, and multipath fading is better modeled as Rician \cite{Rf2}. Commonly encountered Rician communication scenarios include mm-wave massive MIMO communications \cite{Rf3}, and \cite{Rf4} in open air festival. In this paper, we consider crowded massive MIMO communications over Rician fading channels. Each user is assigned a unique pilot-hopping pattern. A joint user identification and LOS component derivation algorithm is proposed, and no threshold is needed to identify the active users. Based on the LOS component, a LOS-only channel estimator and an updated channel estimator are derived.

\begin{figure}[!t]
\centering
\includegraphics [width=3.4 in]{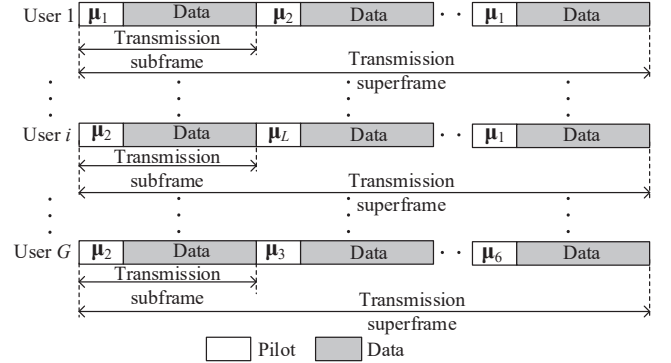}
\caption{Transmission frame structure for crowded massive MIMO communications. (One superframe consists of $U$ subframes. There are $K$ users in all, and $G$ users are active in a specific time slot.)}
\label{fig5a}
\vspace*{-0.15in}
\end{figure}

\section{System Model}
Consider a single cell scenario, where a BS equipped with $M$ antennas ($M\gg 1$) serves $K$ users (or $K$ devices), and each user is equipped with one antenna. A crowded massive communication scenario is taken into account, where the number of users $K$ is much greater than the number of available orthogonal pilot sequences. The goal is to detect the active users and estimate channel impulse responses for the uplink transmission. The proposed algorithm is applicable to both frequency division duplex (FDD) and time division duplex (TDD) systems.

The uplink transmission structure is shown in Fig. 1, where one superframe consists of $U$ subframes. There are $K$ users in all, and $G$ users are active in a specific time slot. The $M\times 1$ channel vector from the $j$-th user to the BS during the $t$-th transmission subframe is \cite{R19}
\begin{equation}
{{\bf{h}}_{t,j}} = \underbrace {{g_j}\sqrt {\frac{{{\kappa _j}}}{{{\kappa _j} + 1}}} {{{\bf{\bar c}}}_j}}_{ \buildrel \Delta \over = {{{\bf{\bar h}}}_j}} + \underbrace {{g_j}\sqrt {\frac{1}{{{\kappa _j} + 1}}} {{{\bf{\tilde c}}}_{t,j}}}_{ \buildrel \Delta \over = {{{\bf{\tilde h}}}_{t,j}}},
\label{eqn1}
\end{equation}
where $g_j$ is the large scale fading factor from the $j$-th user to the BS; ${\kappa _j}$ denotes the Rician factor of the channel from the $j$-th user to the BS; $ \buildrel \Delta \over = $ stands for definition; ${{\bf{\tilde c}}_{t,j}}$ is related to the non-LOS (NLOS) component, each element of which is a complex circularly symmetric Gaussian random variable with zero mean and unit variance, that is ${{\bf{\tilde c}}_{t,j}} \sim CN(0,{{\bf{I}}_M})$; and ${{\bf{\bar c}}_j}$ is related to the LOS component.

When a uniform linear array (ULA) is employed at the BS, ${{\bf{\bar c}}_j}$ can be expressed as \cite{Tse05}
\begin{IEEEeqnarray}{rCl}\label{eqn2}
{{{\bf{\bar c}}}_j} & = & \sqrt M {\bm{\alpha }}({\theta _j}) \nonumber\\
& = & {[1\quad {e^{ - j2\pi d/\lambda \cos ({\theta _j})}} \cdots \quad {e^{ - (M - 1)j2\pi d/\lambda \cos ({\theta _j})}}]^T},
\end{IEEEeqnarray}
where ${\bm {\alpha}}(\cdot)$ is the normalized steering vector; ${\theta _j}$ is the angle of arrival (AOA) of the LOS component from the $j$-th user to the BS; $\lambda$ is the carrier wavelength; and $d$ is the distance between the adjacent antennas of the BS. The channel impulse response ${\bf h}_{t,j}$ remains the same within one transmission subframe. The large scale fading coefficient and the LOS component are assumed to be constant (slowly varying) during one transmission superframe, and the index $t$ is omitted for $g_j$ and ${{\bf{\bar h}}_j}$.

When the BS is equipped with a uniform planar array (UPA), which is in the $yz$-plane with $M_1$ and $N_1$ elements on the $y$ and $z$ axes, respectively, ${\bf{\bar c}}_j$ can be expressed as \cite{Rf5a,Rf6a}
\begin{IEEEeqnarray}{rCl}\label{eqn2c}
{{\bf{\bar c}}_j} & = & \left[ {\begin{array}{*{20}{c}}
{\rm{1}}& \ldots &{{e^{j2\pi d/\lambda (m\sin ({\varphi _j})\sin ({\theta _j}) + n\cos ({\theta _j}))}}}
\end{array}} \right. \\
 &  & \left. {\begin{array}{*{20}{c}}
 \ldots &{{e^{j2\pi d/\lambda ((M_1 - 1)\sin ({\varphi _j})\sin ({\theta _j}) + (N_1 - 1)\cos ({\theta _j}))}}}
\end{array}} \right], \nonumber
\end{IEEEeqnarray}
where ${\varphi _j}$ and ${\theta _j}$ are the azimuth and elevation AOAs of the LOS component from the $j$-th user, respectively. The proposed algorithm is adopted for the case that BS is equipped with a ULA or a UPA. In the following analysis, it is assumed that the BS is equipped with a ULA.

The length of the uplink pilot of each transmission subframe is $L$, and there are $L$ orthogonal pilot sequences (${{\bm{\mu }}_1},{{\bm{\mu }}_1},\ldots,{{\bm{\mu }}_L}$), that is
\begin{equation}
{\bm{\mu }}_k^H{{\bm{\mu }}_j} = \left\{ {\begin{array}{*{20}{l}}
{L,\;\;\;{\kern 1pt} k = j}\\
{0,\;\;\;{\kern 1pt} k \ne j}
\end{array}} \right.,
\label{eqn3}
\end{equation}
where $(\cdot)^H$ stands for conjugate transpose, and the size of ${{\bm{\mu }}_j}$ is $L\times 1$.

Assumed that $L \ll K$ in the crowded massive MIMO communication, and users access the BS sporadically. Thus each user cannot be assigned a unique pilot. In the proposed algorithm, multiple subframes are employed, and the indexes of the pilots in different subframes constitute a pilot hopping sequence. We guarantee that the pilot hopping sequences of different users are different. The pilots for different users in different subframes are assigned according to the pilot hopping sequence.

When the active users transmit pilots in the $t$-th transmission subframe, the received signal of the BS is expressed as
\begin{equation}
{{\bf{Y}}_{t,p}} = \sum\limits_{j = 1}^G {{{\bf{h}}_{t,j}}\sqrt {{p_{p,j}}} {{({{\bm{\mu }}_{I(t,j)}})}^T}}  + {{\bf{W}}_{t,p}},
\label{eqn4}
\end{equation}
where $(\cdot)^T$ stands for transpose; $I(t,j) \in \{ 1,2,...,L\} $ is the index function with respect to the parameters $t$ and $j$; ${{\bm{\mu }}_{I(t,j)}} \in \left\{ {{{\bm{\mu }}_1},{{\bm{\mu }}_2},...,{{\bm{\mu }}_L}} \right\}$ is the pilot sequence employed by the $j$-th user in the $t$-th transmission subframe for channel estimation; ${{\bf{W}}_{t,p}}$ is a complex Gaussian noise matrix, each element of which has a zero mean and variance $\sigma_w^2$; and ${p_{p,j}}$ is the average transmit power of the $j$-th user during the pilot transmission. When the data is transmitted in the $t$-th transmission subframe, the received signal of the BS is given by
\begin{equation}\label{eqn5}
{{\bf{Y}}_{t,u}} = \sum\limits_{j = 1}^G {{{\bf{h}}_{t,j}}\sqrt {{p_{u,j}}} {{\bf{x}}_{t,j}}}  + {{\bf{W}}_{t,u}},
\end{equation}
where ${\bf{x}}_{t,j}$ is the signal vector, whose elements are zero mean and unit variance random variables; $p_{u,j}$ is the average transmit power of the $j$-th user during the data transmission period; and ${{\bf{W}}_{t,u}}$ is a complex Gaussian noise matrix with the same statistical properties as ${{\bf{W}}_{t,p}}$.

\section{Proposed User Identification and Channel Estimation Schemes}
\subsection{The pilot allocation and user identification scheme in \cite{R12}}
In \cite{R12}, the BS performs channel estimation in every transmission subframe (called transmission slot in  \cite{R12}) based on the candidates of the pilot sequences. In the $t$-th transmission subframe, correlating ${{\bf{Y}}_{t,p}}$ with the $l$-th pilot sequence yields
\begin{IEEEeqnarray}{rCl}\label{eqn6}
{{\bf{r}}_{t,l}} & = &  {{\bf{Y}}_{t,p}}{\bm{\mu }}_l^*/(L\sqrt{p_t}) \nonumber\\
& = & \sum\limits_{j = 1}^G {\left( {{{\bf{h}}_{t,j}}\sqrt {{p_t}} {\bm{\mu }}_{I(t,j)}^T{\bm{\mu }}_l^*} \right)}/(L\sqrt{p_t})  + {{\bf{W}}_{t,p}}{\bm{\mu }}_l^*/(L\sqrt{p_t}) \nonumber \\
& = & \sum\limits_{j = 1}^G {{{\bf{h}}_{t,j}} \delta \left( {I(t,j) - l} \right)}  + {{\bf{W}}_{t,p}}{\bm{\mu }}_l^*/(L\sqrt{p_t}),
\end{IEEEeqnarray}
where $(\cdot)^*$ stands for conjugation, $\delta(\cdot)$ is the dirac delta function, and $p_{p,j}=p_{t}$ ($j=1,2,\cdots,G$) is assumed.

Each user is assigned a unique pseudorandom pilot-hopping pattern in \cite{R12} and the BS knows the pilot-hopping pattern of each user in advance. Define ${a_{t,l}} = {\left\| {{{\bf{r}}_{t,l}}} \right\|}$, where ${\left\|  \cdot  \right\|}$ stands for Frobenius norm. With a predefined threshold, the BS determines the pilot sequence employed in each transmission subframe, and identifies active users according to the pseudo-random pilot hopping sequence. However, it is usually difficult to define the threshold in practice. Besides, LOS components were not considered in \cite{R12}. In a Rician fading channel, a LOS component exists between the BS and the user. We exploit this information and design a joint user identification and channel estimation algorithm for crowded massive MIMO communications over Rician fading channels.

\vspace{-0.15in}
\subsection{Proposed algorithm}
The BS applies an angle estimation algorithm within one transmission superframe to derive the LOS AOAs of the active users. The classic multiple signal classification (MUSIC) or estimation of signal parameters via rotational invariance technique (ESPRIT) \cite{R20,R21} can be applied for AOA estimation and the details are not repeated here. The estimated AOAs of the $G$ active users are ${\varphi _1},{\varphi _2}, \ldots ,{\varphi _G}$, respectively, but the BS does not know which AOA corresponds which particular user. The steering vectors corresponding to the estimated AOAs are ${\bm{\alpha }}({\varphi _1}),{\bm{\alpha }}({\varphi _2}), \ldots ,{\bm{\alpha }}({\varphi _G})$.

\subsubsection{Proposed user identification algorithm}
The channel state information (CSI) of active users is contained in ${{\bf{r}}_{t,l}}$. The LOS component related to the LOS AOA is a part of CSI. We project ${{\bf{r}}_{t,l}}$ $(t=1,\cdots,U; l=1,\cdots,L)$ to the steering vector ${\bm{\alpha }}({\varphi _k})$ $(k=1,2,\cdots, G)$ as
\begin{equation}\label{eqn7}
\beta _{t,l}^{(k)} = {\bm{\alpha }}{({\varphi _k})^H}{{\bf{r}}_{t,l}}.
\end{equation}

Define
\begin{equation}\label{eqn8}
\eta _{t}^{(k)} = \arg \mathop {\max }\limits_l \left\{ {\left| {\beta _{t,l}^{(k)}} \right|} \right\},
\end{equation}
and the pattern of the $k$-th steering vector as $\{\eta _{1}^{(k)},\eta _{2}^{(k)}, \ldots ,\eta _{U}^{(k)}\}$.

\Thm If the pattern of the $k$-th steering vector is the same as the pilot-hopping pattern of the $i$-th user, the $i$-th user is an active user and its LOS AOA is $\varphi_k$.

$Proof$: Let the pilot-hopping pattern of the $i$-th user be $\{a_{1,i},a_{2,i},\cdots,a_{U,i}\}$. When the $i$-th user is active, ${\bf{r}}_{1,a_{1,i}}$, ${\bf{r}}_{2,a_{2,i}}$,$\cdots$, ${\bf{r}}_{U,a_{U,i}}$ contain the LOS component of the $i$-th user, which is ${\bar {\bf h}}_i={{g_i}\sqrt {\frac{{{\kappa _i}}}{{{\kappa _i} + 1}}} {\sqrt{M}{{\bm{\alpha}}}(\theta_i)}}$. The component $\left|{\bm{\alpha }}{(\theta_i)^H}{\bar {\bf h}}_j\right|$ is maximized only if $j$ is equal to $i$. The LOS component ${\bar {\bf h}}_i$ is contained in ${\bf{r}}_{1,a_{1,i}}$, ${\bf{r}}_{2,a_{2,i}}$,$\cdots$, ${\bf{r}}_{U,a_{U,i}}$. The pattern of the steering vector ${\bm{\alpha }}(\theta_i)$ is $\{a_{1,i},a_{2,i},\cdots,a_{U,i}\}$, which is the same as the pilot-hopping pattern of the $i$-th user. Therefore, if the pattern of the steering vector ${\bm{\alpha }}(\varphi_k)$ is the same as the pilot-hopping pattern of the $i$-th user, the $i$-th user is an active user and its LOS AOA is $\theta_i=\varphi_k$.

For example, let us assume the following user and hopping pattern pairs: user 1 - $\{1,2,\ldots, 4\}$; user 4 - $\{3,1,\ldots, 4\}$; user 8 - $\{5,2,\ldots, 1\}$. The estimated patterns of the steering vectors ${\bm{\alpha }}(\varphi_1)$, ${\bm{\alpha }}(\varphi_2)$, and ${\bm{\alpha }}(\varphi_3)$ are, respectively,
$\{\eta^{(1)}_1=5,\eta^{(1)}_2=2,\cdots,\eta^{(1)}_U=1\}$, $\{\eta^{(2)}_1=1,\eta^{(2)}_2=2,\cdots,\eta^{(2)}_U=4\}$, $\{\eta^{(3)}_1=3,\eta^{(3)}_2=1,\cdots,\eta^{(3)}_U=4\}$.
Therefore, the BS can identify the active users, that is, the first user, the fourth user, and the eighth user are active. The LOS AOAs corresponding to the first user, the fourth user, and the eighth user are ${\varphi _2}$, ${\varphi _3}$, and ${\varphi _1}$, respectively.

Note that the proposed algorithm detects active users based on the steering vectors of the LOS AOAs, which are related to the LOS components.

\subsubsection{Proposed LOS-only channel estimation algorithm}
Let ${l_{t,j}}$ and ${\varphi _{{k_j}}}$ be corresponding to the estimated pilot index in the $t$-th transmission subframe and the estimated LOS AOA of the $j$-th user, respectively. The average value of $\beta _{t,{l_{t,j}}}^{({k_j})}$ over the $U$ transmission subframes is defined as
\begin{equation}\label{eqn9}
\bar \beta _{(j)}^{({k_j})}{\rm{ = }}\frac{1}{U}\sum\limits_{t = 1}^U {\beta _{t,{l_{t,j}}}^{({k_j})}}.
\end{equation}
Based on ${\varphi _{{k_j}}}$, the LOS component of the $j$-th user can be estimated as
\begin{equation}\label{eqn10}
{{\bf{\hat {\bar h}}}_{j}} = \bar \beta _{(j)}^{({k_j})}{\bm{\alpha }}({\varphi _{{k_j}}}).
\end{equation}
In the LOS-only channel estimation algorithm, the estimated channel is ${{\bf{\hat h}}_{t,j}}={{\bf{\hat {\bar h}}}_{j}} $.

\subsubsection{Proposed updated channel estimation algorithm}
In the updated channel estimation algorithm, the BS firstly applies coherent detection based on the estimated LOS component for each user. Assume that ${p_{p,j}}{\rm{ = }}{p_{u,j}}{\rm{ = }}{p_t}$ for the uplink transmission. In the $t$-th transmission subframe, the detected data of the $j$-th user can be expressed as
\begin{IEEEeqnarray}{rCl}\label{eqn11}
{{{\bf{\hat x}}}_{t,j}} & = & \frac{{{{{\bf{\hat {\bar h}}}}^H}_{j}{{\bf{Y}}_{t,u}}}}{{{{\left| {\bar \beta _{(j)}^{({k_j})}} \right|}^2}\sqrt {{p_t}} }}.
\end{IEEEeqnarray}

Define the following matrix:
\begin{equation}\label{eqn12}
\left\{ \begin{array}{l}
{{\bf{X}}_t} = {\left[ {\begin{array}{*{20}{c}}
{{\bf{x}}_{t,1}^T}&{{\bf{x}}_{t,2}^T}& \cdots &{{\bf{x}}_{t,G}^T}
\end{array}} \right]^T}\\
{{{\bf{\hat X}}}_t} = {\left[ {\begin{array}{*{20}{c}}
{{\bf{\hat x}}_{t,1}^T}&{{\bf{\hat x}}_{t,2}^T}& \cdots &{{\bf{\hat x}}_{t,G}^T}
\end{array}} \right]^T}\\
{{{\bf{\tilde H}}}_t} = \left[ {\begin{array}{*{20}{c}}
{{{{\bf{\tilde h}}}_{t,1}}}&{{{{\bf{\tilde h}}}_{t,2}}}& \ldots &{{{{\bf{\tilde h}}}_{t,G}}}
\end{array}} \right]\\
{\bf{\hat {\bar H}}} = \left[ {\begin{array}{*{20}{c}}
{{{{\bf{\hat {\bar h}}}}_{(1)}}}&{{{{\bf{\hat {\bar h}}}}_{(2)}}}& \ldots &{{{{\bf{\hat {\bar h}}}}_{(G)}}}
\end{array}} \right]
\end{array} \right..
\end{equation}

From the estimated data, the BS can derive the NLOS component. Define
\begin{IEEEeqnarray}{rCl}\label{eqn13}
{{{\bf{\tilde Y}}}_{t,u}} & = & {{\bf{Y}}_{t,u}} - {\bf{\hat {\bar H}}}\sqrt {{p_t}} {{{\bf{\hat X}}}_t} \nonumber\\
& =& \sum\limits_{i = 1}^G {{{\bf{h}}_{t,i}}\sqrt {{p_t}} {{\bf{x}}_{t,i}}}  - \sum\limits_{i = 1}^G {{{{\bf{\hat {\bar h}}}}_{(i)}}\sqrt {{p_t}} {{{\bf{\hat x}}}_{t,i}}}  + {{\bf{W}}_{t,u}} \nonumber\\
& =& \sum\limits_{i = 1}^G {({{{\bf{\bar h}}}_{(i)}} + {{{\bf{\tilde h}}}_{t,i}})\sqrt {{p_t}} {{\bf{x}}_{t,i}}}  - \sum\limits_{i = 1}^G {{{{\bf{\hat {\bar h}}}}_{(i)}}\sqrt {{p_t}} {{{\bf{\hat x}}}_{t,i}}}  + {{\bf{W}}_{t,u}}  \nonumber\\
 &\mathop = \limits^{\kappa,M \to \infty }&  \sum\limits_{i = 1}^G {{{{\bf{\tilde h}}}_{t,i}}\sqrt {{p_t}} {{\bf{x}}_{t,i}}}  + {{\bf{W}}_{t,u}},
\end{IEEEeqnarray}
where the length of ${{\bf{\hat x}}_{t,j}}$ employed for channel estimation is assumed to be $\tau$ $(G<\tau )$. When the Rician factor $\kappa$ and $M$ approach infinity, the estimates of AOA and ${{\bf{x}}_{t,j}}$ will become very accurate. Therefore, the last equation of (\ref{eqn13}) holds.

The NLOS components of the channel from the users to the BS can be estimated by applying minimum mean square error (MMSE) algorithm on the derived data. The estimate of the NLOS components is expressed as
\begin{IEEEeqnarray}{rCl}\label{eqn14}
{{{\bf{\hat {\tilde H}}}}_t} & = & {{{\bf{\tilde Y}}}_{t,u}}{\left( {\sqrt {{p_t}} {{{\bf{\hat X}}}_t}} \right)^H}{\left( {{p_t}{{{\bf{\hat X}}}_t}{{\left( {{{{\bf{\hat X}}}_t}} \right)}^H}{\rm{ + }}\sigma _w^2{\bf{R}}_{v}^{ - 1}} \right)^{{\rm{ - 1}}}} \nonumber\\
&  = &  {{{\bf{\tilde H}}}_t}\sqrt {{p_t}} {{{\bf{\hat X}}}_t}{\left( {\sqrt {{p_t}} {{{\bf{\hat X}}}_t}} \right)^H}{\left( {{p_t}{{{\bf{\hat X}}}_t}{{\left( {{{{\bf{\hat X}}}_t}} \right)}^H}{\rm{ + }}\sigma _w^2{\bf{R}}_{v}^{ - 1}} \right)^{{\rm{ - 1}}}}   \nonumber\\
& &  + {{\bf{W}}_{t,u}}{\left( {\sqrt {{p_t}} {{{\bf{\hat X}}}_t}} \right)^H}{\left( {{p_t}{{{\bf{\hat X}}}_t}{{\left( {{{{\bf{\hat X}}}_t}} \right)}^H}{\rm{ + }}\sigma _w^2{\bf{R}}_{v}^{ - 1}} \right)^{{\rm{ - 1}}}} \nonumber\\
&  = &  \left[ {\begin{array}{*{20}{c}}
{{{{\bf{\tilde h}}}_{t,1}}}&{{{{\bf{\tilde h}}}_{t,2}}}& \ldots &{{{{\bf{\tilde h}}}_{t,G}}}
\end{array}} \right] + {{\bf{\tilde E}}_t},
\end{IEEEeqnarray}
where ${{\bf{\hat {\tilde H}}}_t} = \left[ {\begin{array}{*{20}{c}}
{{{{\bf{\hat {\tilde h}}}}_{t,1}}}&{{{{\bf{\hat {\tilde h}}}}_{t,2}}}& \ldots &{{{{\bf{\hat {\tilde h}}}}_{t,G}}}
\end{array}} \right]$ is the estimate of  ${{\bf{\tilde H}}_t}$, the matrix ${{\bf{R}}_{v}} = E\left[ {{{\bf{v}}_{t,i}}{\bf{v}}_{t,i}^H} \right]$ is related to the statistical properties of the channel and is irrelevant to the index $i$, ${{\bf{v}}_{t,i}} = {\left[ {\begin{array}{*{20}{c}}
{{{\left( {{{{\bf{\tilde h}}}_{t,1}}} \right)}_i}}&{{{\left( {{{{\bf{\tilde h}}}_{t,2}}} \right)}_i}}& \ldots &{{{\left( {{{{\bf{\tilde h}}}_{t,G}}} \right)}_i}}
\end{array}} \right]^T}$, ${\left( {{{{\bf{\tilde h}}}_{t,j}}} \right)_i}$  stands for the $i$-th component of ${{\bf{\tilde h}}_{t,j}}$, and ${{\bf{\tilde E}}_t}=\left[ {\begin{array}{*{20}{c}}
{{{\bf{e}}_{t,1}}}&{{{\bf{e}}_{t,2}}}& \cdots &{{{\bf{e}}_{t,G}}}
\end{array}} \right]$ is the error matrix. Eq. (\ref{eqn14}) shows that the estimation accuracy depends on the accuracy of the estimated data ${{\bf{\hat x}}_{t,j}}$. The updated channel estimate of the $j$-th user is expressed as
\begin{equation}\label{eqn15}
{{\bf{\hat h}}_{t,j}} = {{\bf{\hat {\bar h}}}_j}{\rm{ + }}{{\bf{\hat {\tilde h}}}_{t,j}}.
\end{equation}
The channels of other active users are estimated with the same method in the BS.

The proposed algorithm for the user identification and channel estimation (including the LOS-only channel estimation and the updated channel estimation) depends on the LOS component. Therefore, the proposed algorithm is applicable to the channel with a LOS component, for example, Rician fading channel. It has a good performance when the LOS component dominates the channel. When applied to a channel without a LOS component, for example Rayleigh fading channel, the performance of the proposed algorithm will be worse.

\vspace{-0.2in}
\subsection{Complexity analysis}
The complexity of the proposed scheme mainly comes from the AOA estimation and maximum matching process. The computational complexities of MUSIC AOA estimation is $O(M^3+(M+1)(M-G)r)$ ($r$ is the resolution)  \cite{R21}. The maximum matching process complexity is $O(UGLM)$. The computational complexity of the proposed user identification algorithm is $O(M^3+(M+1)(M-G)r+UGLM)$.

The computational complexities of the proposed scheme, the random access protocol \cite{R12} \cite{Rf6}, and the AMP based scheme \cite{Rf7} are compared in Table I. It can be seen from Table I that the random access protocol [4] has the lowest complexity. The complexity of the proposed scheme is lower than that of the AMP based scheme, when the number of users $K$ or the number of iterations in the AMP based scheme is big.

\begin{table}[!b]
\vspace*{-0.15in}
\caption{Complexity comparison}
\label{ta1}
\centering
\vspace*{-0.10in}
\begin{tabular}{lp{3cm}p{3cm}}
    \hline
   \textbf{Scheme} & \textbf{Complexity}\\ \hline \hline
     Proposed scheme &  $O(M^3+(M+1)(M-G)r+UGLM)$\\ \hline
     Random access protocol \cite{R12} \cite{Rf6}& $O(ULM) $\\ \hline
     AMP based scheme per iteration \cite{Rf7} & $O(KLM)$ \\ \hline
\end{tabular} 
\end{table}

\vspace*{-0.15in}
\section{Simulation Results}
In this section, we show the simulation results to validate the proposed scheme. There are a total of $K=250$ users, and $G$ users are active in each transmission superframe. Every coherent block consists of $T_c=200$ samples, and the pilot length is $L$ ($L\ll K$). One transmission superframe consists of four transmission subframes, that is, one pilot-pattern codeword is transmitted over four continuous transmission subframes. The maximum number of supported users is $L^4$. The large-scale fading coefficients are normalized and uniformly distributed over $[0.1, 1]$. The BS is equipped with a ULA, and the spacing between two adjacent antennas is $d = \lambda/2$. It is assumed that all users have LOS components, and the LOS AOAs of the users are uniformly distributed over $[0, \pi]$. 

\begin{figure}[!t]
\centering
\includegraphics [width=3.5 in]{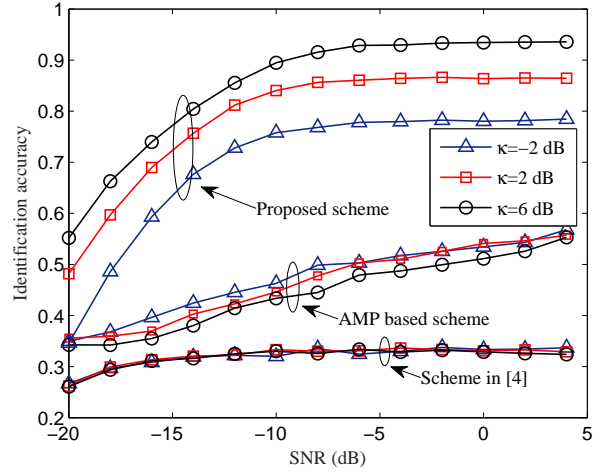}
\caption{Identification accuracy of different schemes in different channel conditions.}
\label{fig2}
\vspace*{-0.15in}
\end{figure}

Signal-to-noise ratio (SNR) is defined as ${\rm{SNR}} = {p_t}/\sigma _w^2$, ${p_{p,j}} = {p_{u,j}} = {p_t}$, and $\kappa_j=\kappa$ ($j=1,2,\cdots,G$). The user identification performances of different schemes are plotted in Fig. \ref{fig2}, where $M=100$, $L=32$, and $G=10$. Fig. \ref{fig2} shows that as the Rician factor decreases, the identification accuracy of the proposed scheme decreases, because the proposed scheme is based on the LOS component.

Compared with the user identification scheme in [4], which does not exploit the information of the LOS components, the performance of the proposed scheme is improved significantly. The proposed user identification algorithm achieves better performance than the AMP based algorithm \cite{Rf7,Rf8}, where the number of iterations should be greater than 3 to achieve convergence and it is set equal to $5$.

\begin{figure}[!t]
\centering
\includegraphics [width=3.5 in]{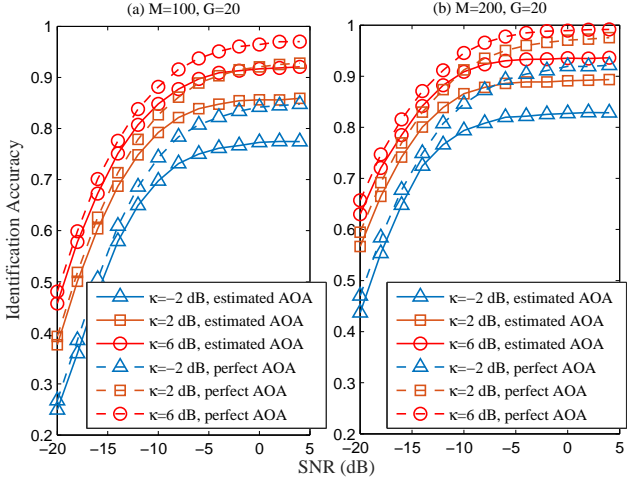}
\caption{Identification accuracy of the proposed algorithm in different settings. (a) $M=100$, $L=16$, $G=20$; (b)$M=200$, $L=16$, $G=20$.}
\label{fig3a}
\vspace*{-0.15in}
\end{figure}

Fig. 3 shows the user identification performance of the proposed scheme with the perfect AOAs and the estimated AOAs in different settings. Compared with the case that assumes perfect AOAs, the accuracy is slightly degraded with estimated AOAs. As the number of BS antennas increases from $100$ to $200$, the performance gap between the case that assumes estimated AOA and the case that assumes perfect AOA increases, which means that the AOA estimation error has a higher impact as the number of BS antennas increases.

\begin{figure}[!t]
\centering
\includegraphics [width=3.5 in]{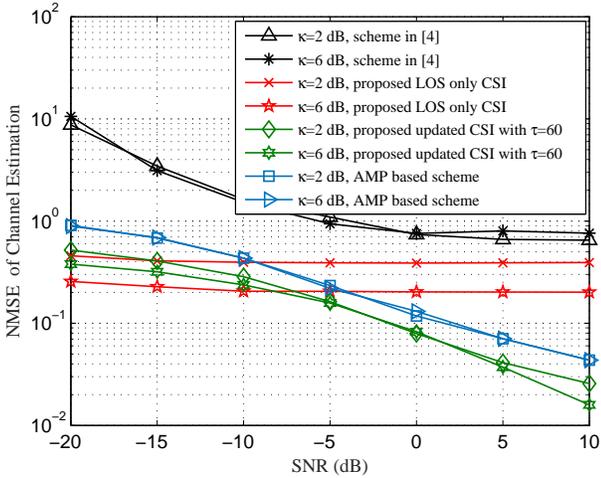}
\caption{Channel estimation performance of different schemes. }
\label{fig3}
\vspace*{-0.15in}
\end{figure}

The normalized mean square error (NMSE) is used to evaluate the performance of the proposed channel estimation schemes, and it is defined as
\begin{equation}\label{eqn40}
{\rm{NMS}}{{\rm{E}}_{t,j}} = \frac{{{{\left\| {{{{\bf{\hat h}}}_{t,j}} - {{\bf{h}}_{t,j}}} \right\|}^2}}}{{{{\left\| {{{\bf{h}}_{t,j}}} \right\|}^2}}}.
\end{equation}
Fig. \ref{fig3} shows the NMSE of estimated channel of different schemes, when $G=10$, $M=100$, and $L=32$. The results reveal that the NMSE of the LOS-only channel estimation scheme is less affected by the SNR, as the estimated channel of this scheme depends only on the LOS component. In the updated channel estimation scheme, 4-ary quadrature amplitude modulation (QAM) is employed during the data transmission, and $\tau=60$. The NMSE performance of the updated scheme is better than the LOS-only based scheme, when SNR is greater than $-10$dB. It is also observed that as the Rician factor $\kappa$ decreases, the channel estimation performance of the proposed algorithms decreases. To maintain an acceptable NMSE value, the Rician factor $\kappa$ cannot be arbitrarily small.

The performance of scheme in [4] is plotted in Fig. 4 for comparison. It can be seen from Fig. 4 that the proposed scheme achieves a better channel estimation performance than the scheme in [4]. The performance of the AMP based scheme \cite{Rf7,Rf8} is also added in Fig. 4. The performance of the proposed updated scheme is better than that of the AMP based scheme. Both the scheme in [4] and the AMP based scheme are less affected by the Rician factor.

\section{Conclusion}
By making use the AOA information, we have proposed a joint user identification and LOS component derivation algorithm, which does not need a threshold that is difficult to obtain in practice, and the identification accuracy is improved significantly. A LOS-only channel estimator and an updated channel estimator have been proposed based on the derived LOS component. The proposed scheme achieves better performance than reference schemes.

\ifCLASSOPTIONcaptionsoff
  \newpage
\fi

\end{document}